\documentclass[letterpaper]{amsart}

\newtheorem{theorem}{Theorem}[section]

\newtheorem{proposition}[theorem]{Proposition}

\theoremstyle{definition}

\theoremstyle{remark}
\newtheorem{remark}[theorem]{Remark}

\usepackage{graphicx}
\usepackage{color}
\usepackage{amsmath}
\usepackage{amsfonts}
\usepackage{amssymb}
\newcommand{\be}{\begin{equation}}
\newcommand{\ee}{\end{equation}}

\newcommand{\dz}{\wedge}

\newcommand{\ba}{\begin{array}}

\newcommand{\ea}{\end{array}}

\newcommand{\beq}{\begin{eqnarray}}

\newcommand{\eeq}{\end{eqnarray}}

\newtheorem{lm}{lemma}

\newtheorem{thee}{theorem}

\newtheorem{proo}{proposition}

\newtheorem{co}{corollary}

\newtheorem{rem}{remark}

\newtheorem{deff}{definition}

\newcommand{\bd}{\begin{deff}}

\newcommand{\ed}{\end{deff}}

\newcommand{\bl}{\begin{lm}}

\newcommand{\el}{\end{lm}}

\newcommand{\bp}{\begin{proo}}

\newcommand{\ep}{\end{proo}}

\newcommand{\bt}{\begin{thee}}

\newcommand{\et}{\end{thee}}

\newcommand{\bc}{\begin{co}}

\newcommand{\ec}{\end{co}}

\newcommand{\brm}{\begin{rem}}

\newcommand{\erm}{\end{rem}}

\newcommand{\der}{{\rm d}}

\usepackage{t1enc}

\newcommand{\newc}{\newcommand}

\let\ccdot\cdot
\def\cdot{\hbox to 2.5pt{\hss$\ccdot$\hss}}

\newc{\aR}{\mbox{\boldmath{$ R$}}}
\newc{\aS}{\mbox{\boldmath{$ S$}}}
\newc{\aT}{\mbox{\boldmath{$ T$}}}
\newc{\aW}{\mbox{\boldmath{$ W$}}}

\newc{\aK}{\mbox{\boldmath{$ K$}}}
\newc{\aL}{\mbox{\boldmath{$ L$}}}

\usepackage{amssymb}
\usepackage{amscd}

\newcommand{\hook}{\raisebox{-0.35ex}{\makebox[0.6em][r]
{\scriptsize $-$}}\hspace{-0.15em}\raisebox{0.25ex}{\makebox[0.4em][l]{\tiny
 $|$}}}

\newc{\obstrn}[2]{B^{#1}_{#2}}

\newcommand{\rpl}                         
{\mbox{$
\begin{picture}(12.7,8)(-.5,-1)
\put(0,0.2){$+$}
\put(4.2,2.8){\oval(8,8)[r]}
\end{picture}$}}

\newcommand{\lpl}                         
{\mbox{$
\begin{picture}(12.7,8)(-.5,-1)
\put(2,0.2){$+$}
\put(6.2,2.8){\oval(8,8)[l]}
\end{picture}$}}

\usepackage{ifthen}

\newcommand{\bbS}{\mathbb{S}}

\newc{\tensor}[1]{#1}
\newc{\Mvariable}[1]{\mbox{#1}}
\newc{\down}[1]{{}_{#1}}
\newc{\up}[1]{{}^{#1}}

\newc{\JulyStrut}{\rule{0mm}{6mm}}
\newc{\midtenPan}{\mbox{\sf S}}
\newc{\midten}{\mbox{\sf T}}
\newc{\midtenEi}{\mbox{\sf U}}
\newc{\ATen}{\mbox{\sf E}}
\newc{\BTen}{\mbox{\sf F}}
\newc{\CTen}{\mbox{\sf G}}

\def\sideremark#1{\ifvmode\leavevmode\fi\vadjust{\vbox to0pt{\vss
 \hbox to 0pt{\hskip\hsize\hskip1em
 \vbox{\hsize3cm\tiny\raggedright\pretolerance10000
 \noindent #1\hfill}\hss}\vbox to8pt{\vfil}\vss}}}%

\numberwithin{equation}{section}

\newcommand{\bma}{\begin{pmatrix}}
\newcommand{\ema}{\end{pmatrix}}

\newcounter{romenumi}
\newcommand{\labelromenumi}{(\roman{romenumi})}

\begin{document}
\title[Twisting type N vacuums with cosmological constant]{Twisting type N vacuums with cosmological constant\\
{\tiny Dedicated to the memory of Jerzy F Pleba\'nski}}
\author{Pawe\l~ Nurowski} 
\address{Instytut Fizyki Teoretycznej,
Uniwersytet Warszawski, ul. Hoza 69, 00-681 Warszawa, Poland}
\email{nurowski@fuw.edu.pl} 
\thanks{This research was supported by the KBN grant 1 P03B 07529}
\date{\today}

\begin{abstract}
We provide the first examples of vacuum metrics with cosmological constant
which have a twisting quadruple principal null direction.
\vskip5pt\centerline{\small\textbf{MSC classification}: 83C15, 83C20, 83C30,
  83F05}\vskip15pt
\end{abstract}
\maketitle
In a joint paper with Jerzy F Pleba\'nski \cite{pleb}, among other things, we provided a maximally
reduced system of vacuum Einstein equations
$$Ric(g)=\Lambda g$$ for \emph{twisting 
type N spacetimes} (see \cite{KSMH}, Ch. 29, for a definition). 
After writing down the reduced equations
we concluded the first section of \cite{pleb} with a remark that it is 
very difficult to find a single solution to the presented equations
when $\Lambda\neq 0$.\\

Actually the problem of finding twisting type N 
vacuums, with the \emph{cosmological constant} $\Lambda=0$ or not, is
one of the hardest in the theory of algebraically special
solutions. If $\Lambda=0$, the only \emph{explicit} solution is
that of Hauser \cite{Hau}. It is \emph{explicit} in the sense that it can be
expressed in terms of hypergeometric functions. Very few 
other twisting type N vacuums with
$\Lambda=0$ are known. Unlike Hauser's solution they
are not that explicit anymore. At best they may be expressed in terms of a \emph{quite complicated 
nonlinear ODE of the third order} \cite{Chin,PlePrzan,Herlt,Lud}. \\

The main difficulty with twisting type N vacuums with $\Lambda=0$ is
that the equations are strongly overdetermined, and it is very difficult to
guess an ansatz which will not lead to the Minkowski metric as the
only solution.\\

Having all this in mind we present the following metric:
\begin{eqnarray}
&&\quad\quad g=\frac{1}{s^2 y^2\cos^2(\frac{r}{2})}\times\label{met1}\\
&&\Big(\tfrac32(\der x^2+\der
y^2)+(\der x+y^3\der u)\big(y\der r+\tfrac13 y^3\cos r\der
u+(2+\tfrac73\cos r)\der x+2\sin r\der y\big)\Big).\nonumber\end{eqnarray}
Here $(x,y,u,r)$, with ranges 
$-\infty<x,u<\infty$, $0<|y|<\infty$, $|r|<\pi$, are 
coordinates of the spacetime
described by $g$; the quantity $s$ is a \emph{real} nonzero constant,
which we regard as a free parameter. 

A short calculation\footnote{Nowadays this can be done by a symbolic computer
  calculation package!} leads to the following remarkable
\begin{proposition}\label{pl}
The metric (\ref{met1}) satisfies the \emph{vacuum Einstein's equations} 
$$Ric(g)=-s^2 g$$ with a \emph{negative} cosmological constant
$\Lambda=-s^2$. 

\noindent
The metric is of \emph{Petrov type N}. In an apropriate null coframe its
\emph{only nonvanishing} 
Weyl spinor coefficient is $$\Psi_4=\frac{14}{3}\frac{s^2}{y^2}{\rm
  e}^{-\frac{ir}{2}}\cos^3\frac{r}{2},\quad\quad{\rm with}\quad\quad\Psi_0=\Psi_1=\Psi_2=\Psi_3=0.$$ 
The vector field
$k=\partial_r$ is tangent to a \emph{twisting} congruence of null and
shearfree geodesics,
which is \emph{alligned} with the \emph{quadruple principal null direction} of the
Weyl tensor for $g$.  
\end{proposition}

Thus the metric (\ref{met1}) provides an example of a \emph{twisting type N
Einstein metric with} negative \emph{cosmological constant}; an example that
we were missing when writing the joint paper \cite{pleb} 
with Pleba\'nski. \\

The following remarks are in order:
\begin{remark}
The twisting congruence of shearfree and null goedesics tangent to the
vector field $k=\partial_r$ of Proposition \ref{pl} corresponds to
a quite nontrivial 3-dimensional CR structure \cite{lhn}. This is
\emph{not} CR  equivalent, even locally, to the Heisenberg group CR 
structure \cite{lhn}, which is the underlying CR structure of the Hauser
solution \cite{amt}. This implies that the solution (\ref{met1}) does
not lead to a Hauser metric by means of any kind of limitting
procedure, such as for example $\Lambda\to 0$. It is quite ironic that
it is the same CR structure that was used in \cite{pleb}, Section 5, 
to construct an example of a metric which satisfies the Bach equations
and which is not conformal to an Einstein metric.  
\end{remark}
\begin{remark}
The metric (\ref{met1}) is not accidental. It appears as one of the
simplest examples in our recent formulation of the twisting vacuum
Einstein equations with cosmological constant \cite{lhn}. In
\cite{lhn} we generalized our earlier results \cite{ln,nurphd} on 4-dimensional
Lorentzian 
spacetimes $({\mathcal M},g)$ satisfying  
the Einstein equations $R_{\mu\nu}=\Phi k_\mu k_\nu$, 
with $k_\mu$ being a vector field tangent to 
a \emph{twisting} congruence of
\emph{null} and \emph{shearfree geodesics}, to the case of a 
\emph{nonvanishing} cosmological constant $\Lambda$. We proved that the 
metric $g$ of a spacetime  $({\mathcal M},g)$ satisfying 
\be
R_{\mu\nu}=\Lambda g_{\mu\nu}+\Phi k_\mu k_\nu,\label{ee}
\ee
with $k$ as above, \emph{factorizes} as
\be
g=\Omega^{-2} \hat{g},\quad\quad \Omega=\cos(\tfrac{r}{2}),\label{met}\ee
where $\hat{g}$ is \emph{periodic} in terms of the null
cooordinate $r$ along $k=\partial_r$. We further showed that if $g$
satisfies (\ref{ee}), then $\mathcal M$ is a \emph{circle bundle}
$\bbS^1\to\mathcal M\to M$ over a 3-dimensional strictly pseudoconvex 
CR manifold $(M,[(\lambda,\mu)])$, and that 
\be \hat{g}=p^2[\mu\bar{\mu}+\lambda(\der
r+W\mu+\bar{W}\bar{\mu}+H\lambda)],\label{cmet}\ee
with
\be
W=i a {\rm e}^{-i r}+b,\quad H=\frac{n}{p^4}{\rm e}^{2i
  r}+\frac{\bar{n}}{p^4}{\rm e}^{-2i r}+q{\rm e}^{i r}+\bar{q}{\rm
  e}^{-i r}+h.\label{w}\ee
Here $\lambda$ (real) and $\mu$ (complex) are 1-forms on $\mathcal M$ such
that $k\hook\lambda=k\hook\mu=0$, $k\hook\der\lambda=k\hook\der\mu=0$, 
$\lambda\dz\mu\dz\bar{\mu}\neq 0$ and, the 
functions $a,b,n,q$ (complex) and $p,h$ (real), all of which are
independent of $r$, satisfy 
\begin{eqnarray}
a&=&c+2\partial\log p\nonumber\\
b&=&ic+2i\partial\log p\nonumber\\
q&=&\frac{3n+\bar{n}}{p^4}+\frac23\Lambda p^2+\frac{2\partial p\bar{\partial}p-p(\partial\bar{\partial}p+\bar{\partial}\partial
  p)}{2p^2}-\frac{i}{2}\partial_0\log
  p-\bar{\partial}c\label{q}\\
h&=&3\frac{n+\bar{n}}{p^4}+2\Lambda p^2+\frac{2\partial
  p\bar{\partial}p-p(\partial\bar{\partial}p+\bar{\partial}\partial
  p)}{p^2}-\bar{\partial}c-\partial\bar{c}.\nonumber
\end{eqnarray}
Here the $r$-independent complex function $c$ is defined via
\begin{eqnarray}
&&\der\mu=0,\quad\quad\quad\der\bar{\mu}=0,\nonumber\\
\der\lambda&=&i\mu\dz\bar{\mu}+(c\mu+\bar{c}\bar{\mu})\dz\lambda,\label{c}
\end{eqnarray}
and the operators $(\partial_0,\partial,\bar{\partial})$ are 
vector fields on $M$, which
are algebraic dual to the coframe $(\lambda,\mu,\bar{\mu})$ on
the CR manifold $M$. The only unknowns, $n$ and $p$ satisfy
the following system of PDEs on $M$:
\begin{eqnarray}
&&\partial n+3cn=0,\label{m}\\
&&[~\partial\bar{\partial}+\bar{\partial}\partial
   +\bar{c}\partial+c\bar{\partial}+\tfrac12c\bar{c}+\tfrac34(\partial\bar{c}+\bar{\partial}
    c)~]p=\frac{n+\bar{n}}{p^3}+\tfrac23\Lambda p^3.\label{p}
\end{eqnarray}    
Note that the above functions $a,b,c,h,n,p,q$ define a \emph{vacuum}
metric $g$ via (\ref{met}) iff, in addition to equations
(\ref{m})-(\ref{p}), the functions
$n$ and $p$ satisfy also the equation $\Phi=0$. This is quite nasty 
nonlinear PDE relating $n,p$ and $c$.\\

In this context our solution (\ref{met1}) is very simple. It is given by:
\begin{eqnarray*}
&&\lambda=\frac{2y^4}{3}(\der u+\frac{\der x}{y^3}),\quad\quad\mu=\der
x+i\der y,\quad\quad c=-\frac{2i}{y},\\
&&n=0,\quad\quad\quad\quad\quad\quad\quad\quad p=\frac{\sqrt{3}}{2 s
  y},\quad\quad\quad\quad s^2=-\Lambda,
\end{eqnarray*}
as can be easily checked by a direct calculation using definitions
(\ref{met})-(\ref{c}). 
It is a miracle that in \emph{addition} to the pure radiation Einstein equations
(\ref{m})-(\ref{p}), this solution satisfies the \emph{vaccum} condition
$\Phi= 0$, and simultaneously the \emph{type N} conditions $\Psi_0=\Psi_1=\Psi_2=\Psi_3=0$,
$\Psi_4\neq 0$. This shows that the formulation of the
twisting Einstein equations described in \cite{lhn,ln,nurphd} has an
unexplored power. 
\end{remark}

\begin{remark}
Actually the solution (\ref{met1}) is only the simplest one from a larger
class of twisting type N vacuums with cosmological constant 
contained in (\ref{met})-(\ref{p}).
These solutions are obtained by taking
\begin{eqnarray}
&&\lambda=-\frac{2}{f'(y)}\Big(\der u+f(y)\der x\Big),\quad\quad\mu=\der
x+i\der y,\quad\quad c=\frac{if''(y)}{2f'(y)},\label{sol1}\\
&&n=0,\quad\quad\quad\quad\quad\quad\quad\quad p=p(y),\nonumber
\end{eqnarray}
with real functions $f=f(y)$ and $p=p(y)$. With this choice 
the twisting type N vacuum Einstein equations with cosmological
constant are equivalent to the following system of two 3rd
order ODEs:
\begin{eqnarray}
\tfrac94 pf'f'''+3p'f'f''-3p{f''}^2-3p''{f'}^2+4\Lambda p^3 {f'}^2=0,\label{sol2}
\end{eqnarray}
\begin{eqnarray*}
&&3pf'\Big(3{f'}^2p'''+2f''(9p'f''-5f'p'')\Big)+3{f'}^2p'(5f'p''-14p'f'')-12p^2{f''}^3=\\&&152\Lambda
  p^3{f'}^2(pf''-2p'f').
\end{eqnarray*}
It can be checked by direct calculation that every solution $p=p(y)$,
$f=f(y)$ of these two equations, after being inserted into (\ref{sol1})
and (\ref{met})-(\ref{q}), gives a metric $g$, which satisfies
Einstein's vacuum equations $Ric(g)=\Lambda g$ and is of Petrov type
N, having $k=\partial_r$ tangent to a twisting
congruence of null geodesics. The only nonvanishing Weyl spin coefficient
for these metrics is
\begin{eqnarray*}
\Psi_4=-\frac{2{\rm
    e}^{-\frac{ir}{2}}\cos^3\frac{r}{2}}{27p^2{f'}^2}~\Lambda~
\Big(8\Lambda p^4{f'}^2+27{f'}^2{p'}^2+12p^2{f''}^2+3pf'(f'p''-13p'f'')\Big).
\end{eqnarray*}
Equations (\ref{sol2}) admit nontrivial
solutions. The simplest of them leads to our metric (\ref{met1}). Note
that if $\Lambda\to 0$, then also the Weyl coefficient $\Psi_4\to 0$. Hence the
metric $g$ becomes flat. This shows that none of the
solutions (\ref{sol2}) can be obtained as a $\Lambda$-deformation
of the Hauser solution.
\end{remark}
\begin{remark}
Since, as discused in \cite{nuhi}, every metric
(\ref{met})-(\ref{p}) is relevant for Penrose's `before the big bang'
argument \cite{pen}, the type N metrics corresponding to solutions of 
(\ref{sol2}) constitute a nice set of explicit examples in which
Penrose's ideas can be tested.  
\end{remark}

\end{document}